\documentclass[prb, twocolumn, a4paper, superscriptaddress]{revtex4}
\usepackage{natbib}
\usepackage{dcolumn}
\usepackage{setspace}
\usepackage{amsmath}
\usepackage{graphicx}
\usepackage{verbatim}
\usepackage{color}
\usepackage{subfigure}
\usepackage{hyperref}
\usepackage{textcomp}
\usepackage{gensymb}
\usepackage{comment}
\usepackage{siunitx}

\makeatletter

\newcommand{\Rmnum}[1]{\expandafter\@slowromancap\romannumeral #1@}
\makeatother

\begin{document}
\title{Origin of skyrmion lattice phase splitting in Zn-substituted Cu$_{2}$OSeO$_{3}$}
\author{A. \v{S}tefan\v{c}i\v{c}}
\email{A.Stefancic@warwick.ac.uk}
\affiliation{University of Warwick, Department of Physics, Coventry, CV4 7AL, United Kingdom}
\author{S. Moody}
\affiliation{Durham University, Department of Physics, South Road, Durham, DH1 3LE, United Kingdom}
\author{T.J. Hicken}
\affiliation{Durham University, Department of Physics, South Road, Durham, DH1 3LE, United Kingdom}
\author{T.M. Birch}
\affiliation{Durham University, Department of Physics, South Road, Durham, DH1 3LE, United Kingdom}
\author{G. Balakrishnan}
\affiliation{University of Warwick, Department of Physics, Coventry, CV4 7AL, United Kingdom}
\author{S.A. Barnett}
\affiliation{Diamond Light Source, Harwell Science and Innovation Campus, Didcot OX11 0DE, United Kingdom}
\author{M. Crisanti}
\affiliation{University of Warwick, Department of Physics, Coventry,
  CV4 7AL, United Kingdom}
\affiliation{Institut Laue-Langevin,  Large Scale Structures Group,  71 avenue des Martyrs
CS 20156, 38042, Grenoble, Cedex 9,  France}
\author{J.S.O. Evans}
\affiliation{Department of Chemistry, Durham University, South Road, Durham, DH1 3LE, United Kingdom}
\author{S.J.R. Holt}
\affiliation{Durham University, Department of Physics, South Road, Durham, DH1 3LE, United Kingdom}
\author{K.J.A. Franke}
\author{P.D. Hatton}
\author{B.M. Huddart}
\affiliation{Durham University, Department of Physics, South Road, Durham, DH1 3LE, United Kingdom}
\author{M.R. Lees}
\affiliation{University of Warwick, Department of Physics, Coventry, CV4 7AL, United Kingdom}
\author{F.L. Pratt }
\affiliation{ISIS Facility, STFC Rutherford Appleton Laboratory, Chilton Didcot, Oxfordshire, OX11~0QX, United Kingdom}
\author{C.C. Tang}
\affiliation{Diamond Light Source, Harwell Science and Innovation Campus, Didcot OX11 0DE, United Kingdom}
\author{M.N. Wilson}
\affiliation{Durham University, Department of Physics, South Road, Durham, DH1 3LE, United Kingdom}
\author{F. Xiao}
\affiliation{Laboratory for Neutron Scattering, Paul Scherrer Institut, CH-5232 Villigen PSI, Switzerland}
\affiliation{Department of Chemistry and Biochemistry, University of Bern, CH-3012 Bern, Switzerland}
\author{T. Lancaster}
\email{tom.lancaster@durham.ac.uk}
\affiliation{Durham University, Department of Physics, South Road, Durham, DH1 3LE, United Kingdom}

\date{\today}
\begin{abstract}
We present an investigation into the structural and magnetic properties of Zn-substituted Cu$_{2}$OSeO$_{3}$, a system in which the skyrmion lattice  (SkL) phase in the magnetic field-temperature phase diagram was previously seen to split as a function of increasing Zn concentration. We find that splitting of the SkL is only observed in polycrystalline samples and reflects the occurrence of several coexisting phases
with different Zn content, each distinguished by different magnetic behaviour. No such multiphase behaviour is observed in single crystal samples.  
\end{abstract}
\maketitle

%%%%%%%%%%%%%%%%%%%%%%%%%%%%%%%%%%%%%%%%%%%%%%%%%%%%%%%%%%%%%%%%%%%%%%%%%%%%%%%%%%%%%%%%%%%%%%%
There has been considerable recent interest in the synthesis and understanding of  materials hosting topological phases, owing to their exotic physics and potential for applications. Skyrmions are nano-sized, topologically-protected magnetic spin textures, which are promising candidates for energy-efficient, high-density storage devices \cite{liu2016skyrmions, nagaosa2013topological, fert2013skyrmions}.
Skyrmions are found in a number of systems including MnSi\cite{muhlbauer2009skyrmion, tonomura2012real}, $\text{Fe}_{1-x}\text{Co}_x\text{Si}$ \cite{munzer2010skyrmion, yu2010real}, FeGe \cite{yu2011near}, $\beta$-Mn type Co-Zn-Mn alloys \cite{tokunaga2015new}, $\text{Cu}_{2}\text{OSeO}_{3}$ \cite{seki2012observation}, $\text{GaV}_4\text{S}_8$ \cite{kezsmarki2015neel}, and $\text{GaV}_4\text{Se}_8$ \cite{fujima2017thermodynamically}. 
The discovery of skyrmions in the multiferroic insulator, $\text{Cu}_{2}\text{OSeO}_{3}$ represents an important milestone, because of the possibility of manipulating them with an external electric field.\cite{seki2012observation}
%In fact, $\text{Cu}_{2}\text{OSeO}_{3}$ is arguably the most investigated skyrmion hosting material thus far.
Despite an intensive search, the number of known skyrmion-hosting materials still remains relatively small, owing to the very specific structural properties (absence of inversion centre) and magnetic properties (ferromagnetism and presence of Dzyaloshinskii-Moriya interactions) needed to promote the formation of magnetic skyrmions.
An alternative route to expand the number of skyrmionic materials is to utilise a chemical doping/substitution strategy in known skyrmion-hosting systems. 
This approach has recently been adopted in $\text{Cu}_{2}\text{OSeO}_{3}$, where Zn and Ni doping led to observation of splitting of the skyrmion lattice (SkL) phase into two distinct pockets,\cite{wu2015unexpected} and an expansion of the SkL phase \cite{chandrasekhar2016effects} respectively. 

$\text{Cu}_{2}\text{OSeO}_{3}$ crystallises in the non-centrosymmetric cubic $P2_1 3$ space group. 
 Two crystallographically inequivalent $\text{Cu}^{2+}$ cations in  trigonal bipyramidal and  square pyramidal  coordination geometry
(designated as $\text{Cu}^{\mid}$ and $\text{Cu}^{\parallel}$ respectively)
are present in a ratio $1:3$. 
These have spins $S = 1/2$ pointing in opposite directions  forming a ferrimagnetic lattice \cite{bos2008magnetoelectric, belesi2010ferrimagnetism, maisuradze2011mu}.
Magnetic interactions between spins are  mediated through oxygen atoms via ferromagnetic and antiferromagnetic superexchange interactions \cite{yang2012strong, vzivkovic2012two}. 
As reported by  Wu \textit{et al.}, \cite{wu2015unexpected} the magnetic moment of $\text{Cu}_{2}\text{OSeO}_{3}$ monotonically decreases with increasing  Zn-substitution  levels, which they  interpret in terms of the site-specific substitution of $\text{Cu}^{2+}$ cation at the $\text{Cu}^{\parallel}$ site with non-magnetic $\text{Zn}^{2+}$. 
This is accompanied by a splitting of the SkL phase observed in pristine $\text{Cu}_{2}\text{OSeO}_{3}$ into two distinct pockets.

In order to understand the effect of substitution and to confirm the origin of the SkL splitting in this system, investigations on high-quality single crystals are essential. 
We have
 produced single crystals of Zn-substituted $\text{Cu}_{2}\text{OSeO}_{3}$ for several substitution levels of Zn, allowing us to compare their behavior to that of high-purity polycrystalline materials. We find that the splitting of the SkL phase reflects the presence of multiple structural phases in polycrystalline samples. In contrast, phase-pure single crystal samples do not show such behaviour. 

%%%%%%%%%%%%%%%%%%%%%%%%%%%%%%%%%%%%%%%%%%%%%%%%%%%%%%%%%%%%%%%%%%%%%%%%%%%%%%%%%%%%%%%%%%%%%%%

By utilising the chemical vapour phase transport technique, single crystals of Zn-substituted $\text{Cu}_{2}\text{OSeO}_{3}$, with nominal $2,\ 8,$ and $12\%$ Zn content, were grown. 
Energy-dispersive X-ray spectroscopy (EDX), X-ray fluorescence (XRF) and  inductively coupled plasma mass spectroscopy (ICP MS) measurements on single crystals revealed the Zn-concentrations given in Table~\ref{table}. We use the average values of Zn substitution ($0.5, 1.8$ and $2.4\%$) below. 

\begin{table}
\begin{tabular}{c|c|c|c|c|c}
\hline\hline
Nominal & ICP MS & EDX & XRF&Average & Polycrystal\\
\hline
2 &	0.9 &	0.52(9) &	0.12(1) & 0.5 & 2.0\\
8 &	1.9 &	1.71(4) &	1.71(1) & 1.8 & 6.4\\
12 &	2.4 &	2.51(3) &	2.26(1) & 2.4 &10.5\\
\hline \hline
\end{tabular}
\caption{Percentage Zn concentration in (Cu$_{1-x}$Zn$_{x}$)$_{2}$OSeO$_{3}$ single crystals determined by ICP MS, EDX and XRF. Zn concentration in corresponding polycrystalline samples is given in the right-most column.\label{table} }
\end{table}

\begin{figure}
\includegraphics[width=\linewidth]{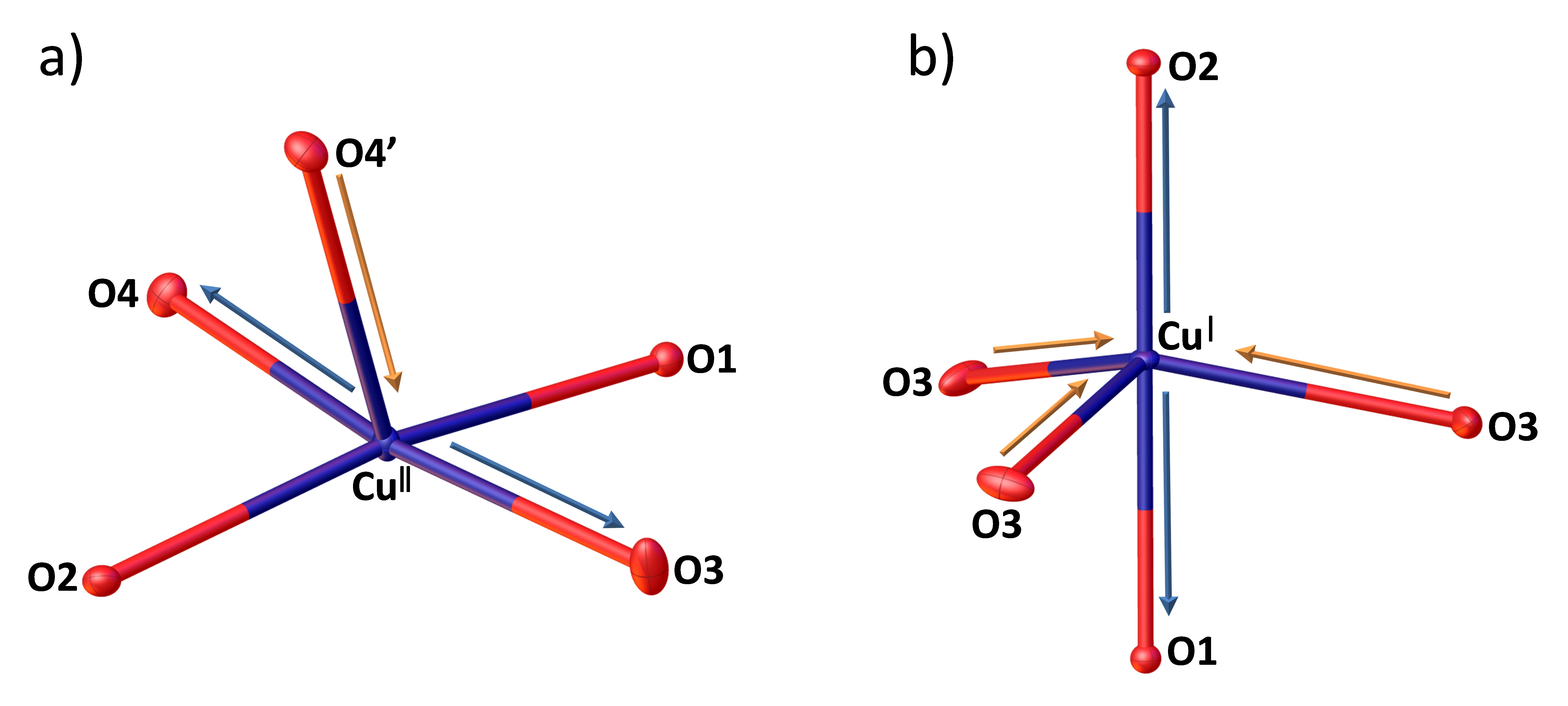}
\caption{Structural model for 2.4\% Zn-substituted Cu$_{2}$OSeO$_{3}$. Coordination environment of (a) the Cu$^{\parallel}$ site (square pyramidal) and (b) the Cu$^{\mid}$ site (trigonal bipyramidal). Arrows indicate elongation and contraction of bond distances compared to pristine  Cu$_{2}$OSeO$_{3}$.
\label{Fig:ales1}}
\end{figure}

The single crystal samples were investigated on the high-resolution I19 beamline at the Diamond Light Source. The $a$ lattice parameter, extracted from 
data collected at the Zr K-edge (17.9976~keV), was found to monotonically increase with increasing Zn substitution levels, as expected when  substituting Cu$^{2+}$ with the slightly larger Zn$^{2+}$ cations (with atomic radii of $0.65$ \AA \ and $0.68$ \AA \ for five coordinate  Cu$^{2+}$/Zn$^{2+}$ respectively \cite{effenberger1986kristallstrukturen, shannon1976revised}). A close comparison of the structural models  for pristine Cu$_{2}$OSeO$_{3}$ single crystals and 2.4\% Zn-substituted Cu$_{2}$OSeO$_{3}$ revealed a very subtle increase of the short Cu$^{\mid}$-O1 and Cu$^{\parallel}$-O2 bond distances of 0.004(2) and 0.002(2)~\AA\ respectively [Fig.~\ref{Fig:ales1}]. This leads to an increase of Cu$^{\mid}$-Cu$^{\parallel}$ distances, while the Cu$^{\parallel}$-Cu$^{\parallel}$ distances remain the same\cite{supplemental}. 

Polycrystalline materials were synthesised as reported in the Supplemental Material \cite{supplemental}.
The phase-purity and lattice parameters were investigated using high-resolution powder x-ray  diffraction (PXRD) at the I11 beamline, Diamond Light Source,  
with  diffraction patterns for  each Zn-substituted polycrystalline samples measured at 15.02~keV. A simple axial model was used to  describe the peak asymmetry at low angles arising from axial divergence of the beam and  a pseudo-Voigt function to  describe the Cu$_{2}$OSeO$_{3}$ peak shapes. 

\begin{figure}
\includegraphics[width=\linewidth]{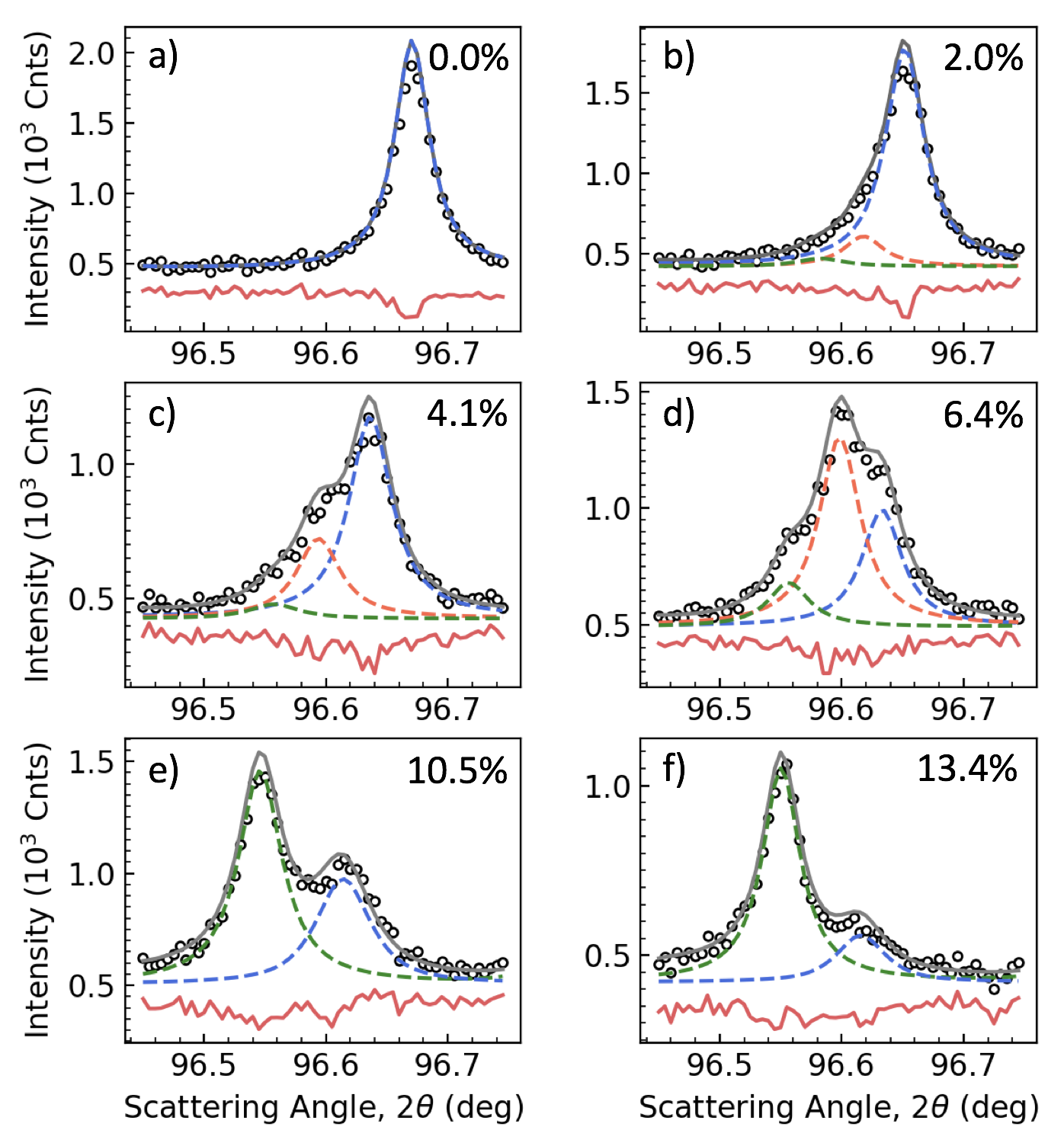}
\caption{
Left inset: high angle part of the pattern. {\it Right inset}:, showing a single Cu$_{2}$OSeO$_{3}$ phase is sufficient to model the data.
 (a)-(f) Powder X-ray diffraction peak at 96.65$^{\circ}$ for polycrystalline samples with different Zn concentrations, demonstrating that the splitting can  be modelled by multiple phases. }
\label{Fig:max3}
\end{figure}

In the unsubstituted material, a single phase of Cu$_{2}$OSeO$_{3}$ is sufficient to model the diffraction pattern, with small residuals. However, for the samples with increased substitution, all of the peaks in the diffraction pattern split, as shown in Figs.~\ref{Fig:max3}(a)-(f). 
Refinements revealed that this splitting could only be effectively modelled by the inclusion of two or three distinct Cu$_{2}$OSeO$_{3}$
phases with differing lattice parameters, suggesting that the polycrystalline materials are multiphase under the synthetic conditions used. 

\begin{figure}
\includegraphics[width=\columnwidth]{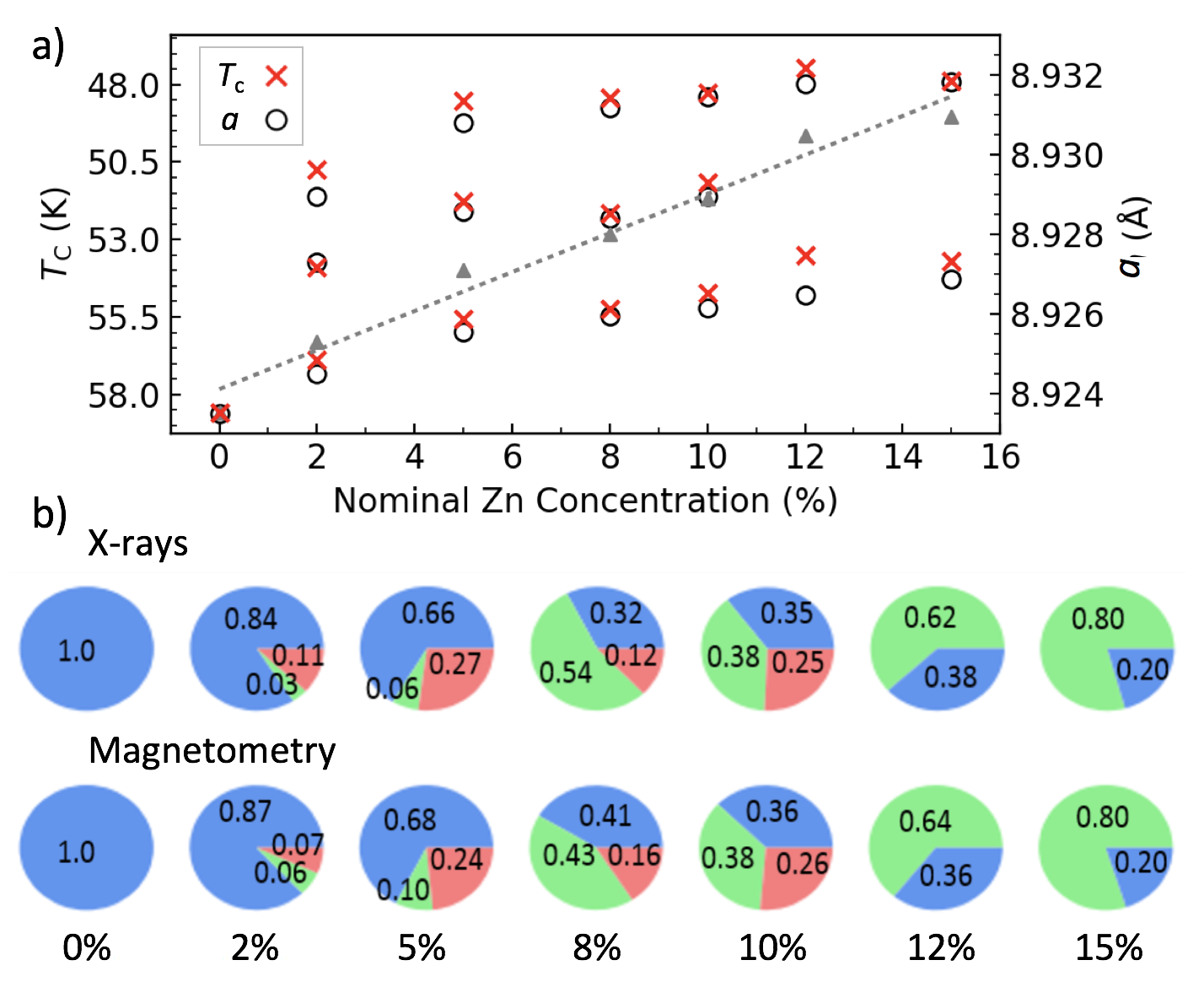}
\caption{(a) Critical temperatures $T_{\mathrm{c}}$ (crosses) and lattice parameter $a$ (circles) for each Cu$_{2}$OSeO$_{3}$ phase in polycrystalline samples,  as a function of nominal Zn-substitution. (Triangles show the average lattice parameter $a$.) (b) Volume fractions of each phase.}
\label{Fig:max4}
\end{figure}

Based on our refinements, the lattice parameter $a$ and cell volume of each phase increases monotonically with increasing Zn-substitution levels  (Fig.~\ref{Fig:max4}), consistent with the behaviour of the single crystals.
Using our PXRD measurements, the estimated Zn-substitution levels in polycrystalline samples are taken as 0, 2, 4.1, 6.4, 7.9 10.5 and $13.4\%$,  for the nominal starting compositions of 0, 2, 5, 8, 10, 12 and $15\%$, respectively.\cite{supplemental}
The substitution level of Cu$^{2+}$ with Zn$^{2+}$ is higher at low nominal substitution values (up to $5\%$ Zn), while above this level it starts to decrease.
This decrease is reflected in the amount of unreacted CuO detected with powder diffraction, increasing from $1\%$ in the sample with $6.4\%$ Zn up to $5\%$ in the sample with $13.4\%$ Zn-substitution level. 
The crystal structure of $\text{Zn}_{2}\text{OSeO}_{3}$, if it exists, is not known, therefore the increase of unreacted CuO with increasing Zn-substitution levels can be related to the stability and Zn$^{2+}$ uptake ability of the $\text{Cu}_{2}\text{OSeO}_{3}$  crystal structure.
In addition, up to $3\%$ of a $\text{Zn}_{2}\text{SiO}_{4}$ impurity phase was observed in samples with Zn-substitution levels higher than $4\%$, indicating that ZnO reacts with silica tubes (although we would normally  expect this reaction to occur at higher temperatures of $> 1000~^{\circ}\text{C}$ \cite{bunting1930phase}).  
The estimated Zn concentrations are given in Table~\ref{table} for comparison with the single crystal samples. We note that the Zn uptake is far smaller in the single crystal samples, compared to the polycrystalline analogues. 

\begin{figure}
\includegraphics[width=\linewidth]{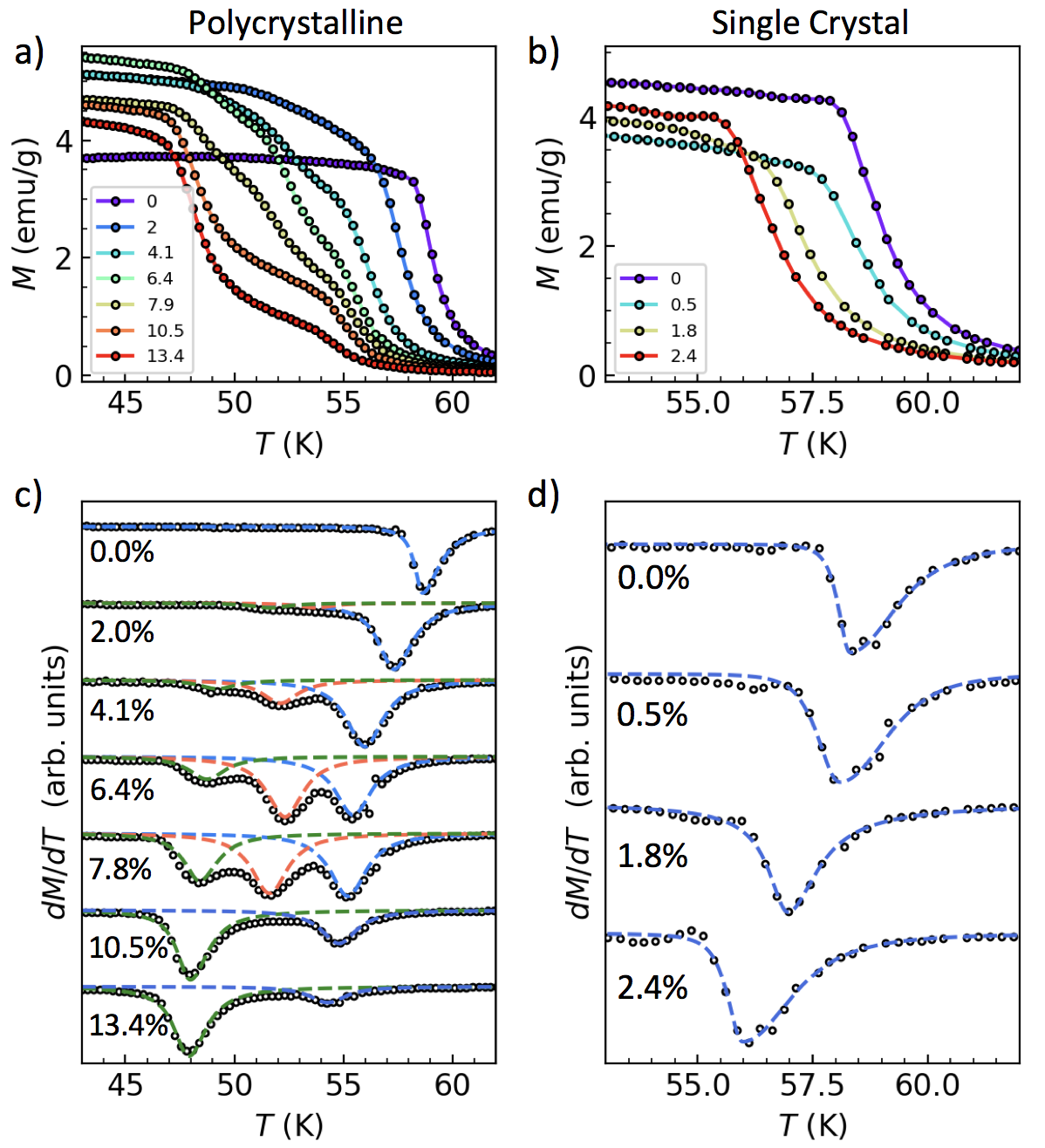}
\caption{(a) Magnetization $M$ vs\ $T$ for polycrystalline samples,  measured in a magnetic field of 25~mT. (b) the same for single crystal samples. (c) The derivative of the $M$ vs\ $T$ polycrystalline data, revealing multiple magnetic transitions. (d) The derivative of the $M$ vs\ $T$ single crystal data, showing a single transition.}
\label{Fig:max2}
\end{figure}

To characterize the magnetic behavior of all samples, 
DC magnetic susceptibility measurements were carried out using zero-field-cooled (ZFC) and field-cooled (FC) protocols. Single crystals were aligned with the applied magnetic field directed along the [111] direction. The resulting magnetisation curves are shown in Figs.~\ref{Fig:max2}(a) and (b).
In single crystals the magnetic transition temperature $T_{\mathrm{c}}$ is found to decrease with increasing Zn content.
However, the decrease is considerably smaller in single crystals compared to polycrystalline materials, consistent with the relatively low percentage of Zn$^{2+}$ inclusion in the $\text{Cu}_{2}\text{OSeO}_{3}$ crystal structure.
 The step-like nature of the data for the polycrystalline material suggests that there are multiple magnetic transitions occurring. 
This is more clearly seen in the derivative of these data [Fig.~\ref{Fig:max2}(c)],  which show multiple peaks, particularly for higher Zn substitution, indicating the presence of several magnetic phases with different values of $T_{\mathrm{c}}$. (This may be contrasted with analogous measurements on single crystals [Fig.~\ref{Fig:max2}(d)] showing a single transition.) 
By fitting the peaks with asymmetric Lorentzian and Split Pearson VII functions for the polycrystalline and single crystal datasets respectively, a critical temperature $T_{\mathrm{c}}$ and volume fraction for each distinct phase was determined, as summarised in Fig.~\ref{Fig:max4}.

\begin{figure}[t]
\includegraphics[width=\linewidth]{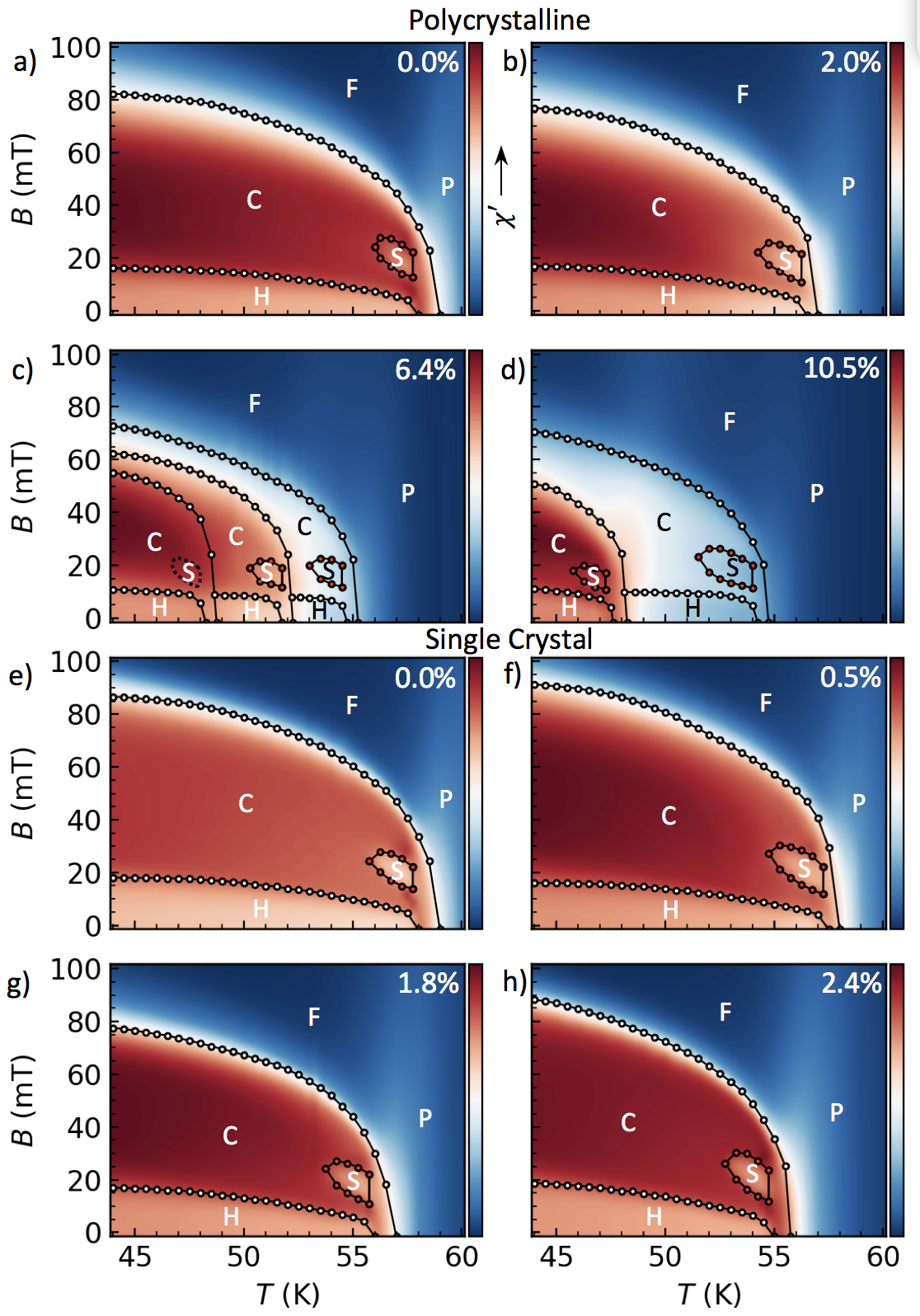}
\caption{ (a)-(d) Real component of AC susceptibility $\chi'$  as a function of field $B$ and temperature $T$ for polycrystalline samples, showing the paramagnetic (P), ferrimagnetic (F), conical (C), helical (H) and skyrmion lattice (S) phases. (e)-(h). The same, but for single crystal samples.}
\label{Fig:max1}
\end{figure}

The existence of skyrmions, along with the size and potential splitting of the skyrmion pocket, were investigated using AC magnetic susceptibility.
The resulting magnetic phase diagrams 
are presented in Fig.~\ref{Fig:max1},
where we show the real component $\chi'$ as a function of applied field and temperature, which provides a good indication of the helical, conical and skyrmion structures.
The phase diagrams for polycrystalline samples closely resemble  the data from Wu {\it et al.}, with $T_{\mathrm{c}}$ of the sample decreasing with increasing Zn substitution, while the skyrmion region seemingly splits into two (or even three in the case of the 6.4\% substituted sample) distinct pockets. [This is also seen in the imaginary component of the susceptibility $\chi'' $\cite{supplemental}.] 
Data measured on  single crystals [Fig.~\ref{Fig:max1}(e-h)],
show a skyrmion phase in each of the samples.
The skyrmion pockets in single crystals with $0.5, 1.8$ and $2.4\%$ Zn-substitution levels stretch from $54.75$ to $57.25$ K, $53.5$ to $56$ K and $52.75$ to $55$ K respectively. Although the  pockets  shift to lower temperatures with increasing Zn concentration, 
their widths remain roughly constant (between $2.25$ and $2.5$~K), in contrast to the skyrmion pockets detected in polycrystalline materials.
The single crystals with $2.4\%$ Zn-substitution level were also measured with $B \mid\mid$ to $[110]$ and the only noticeable differences are in the shape and size of the skyrmion pocket as reported in the literature \cite{adams2012long}. 
For $B \mid\mid$ to $[111]$, the skyrmion pocket can be found in the temperature range between $52.75$ and $55.0$ K, while for $B \mid\mid$  $[110]$ the pocket exists from $53.25$ to $55.0$ K.  

Our analysis of the magnetization measurements suggests the presence of multiple magnetic phases in the polycrystalline samples. To establish whether these are intrinsic to the bulk of each material, we carried out longitudinal field (LF) muon-spin relaxation ($\mu$SR) measurements, since the muon is a sensitive local probe of bulk magnetism. 
\begin{figure}[tb]
	\centering
	\includegraphics[width=0.8\linewidth]{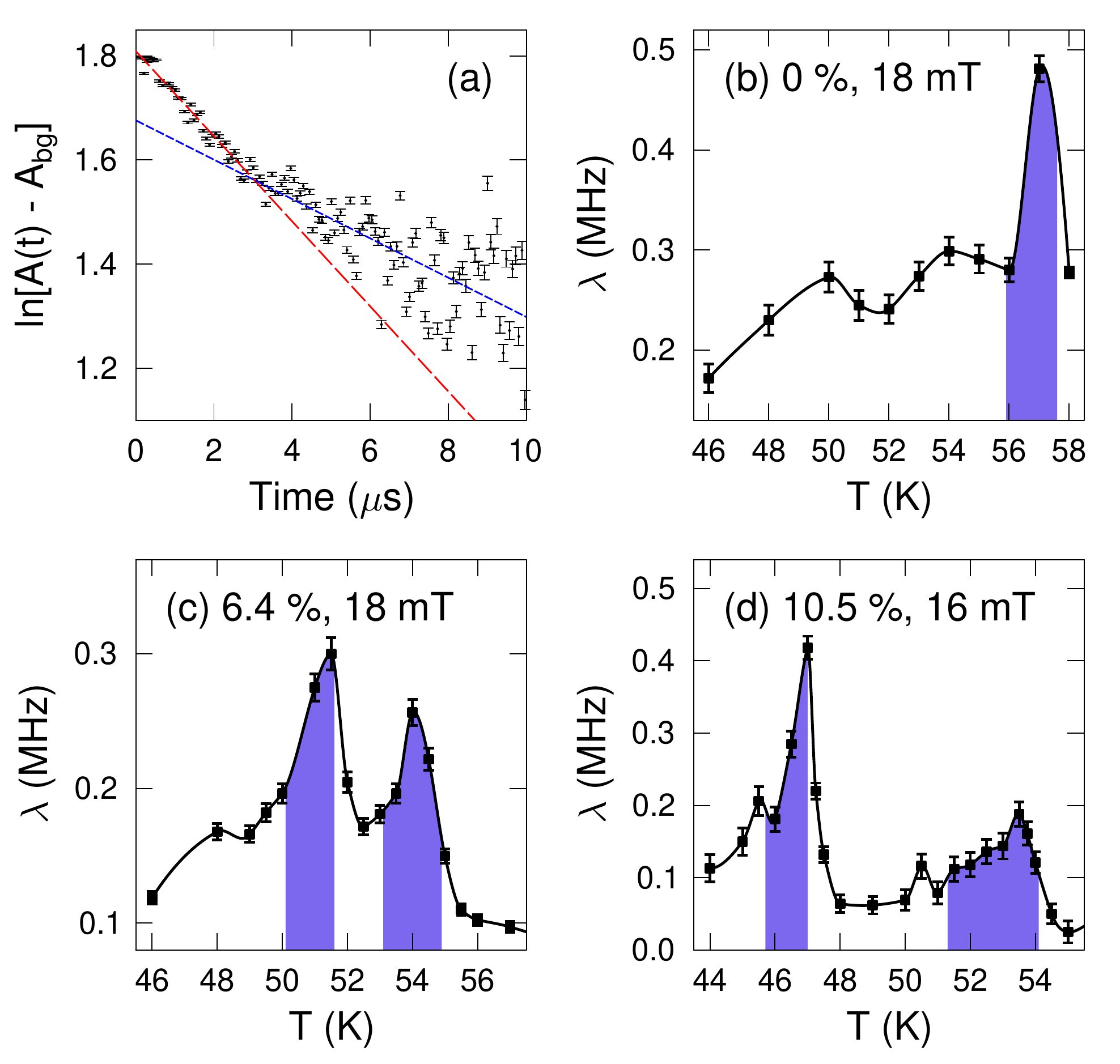}
	\caption{
(a) LF $\mu$SR spectra for 6.4\% Zn-substituted Cu$_2$OSeO$_3$ measured at 54~K in  18~mT  on a  logarithmic scale with the background subtracted. Lines are guides to the eye. Relaxation rate $\lambda$ for LF $\mu$SR measurements of (b) pristine, (c) 6.4\% and (d) 10.5\% Zn-substituted Cu$_2$OSeO$_3$.
Shaded regions indicate the location of the SkL phase derived from AC susceptibility.}
	\label{fig:lambda}
\end{figure}
In order to parametrise the spectra across the measured temperature regime,
they were fitted to an exponential function, with the resulting values of relaxation rate $\lambda$ shown in Fig.~\ref{fig:lambda}.
We find  $\lambda$ to be significantly larger inside the SkL phase than  outside it.
Notably, we observe peaks in $\lambda$ coinciding with the SkL phases which are significantly broader than the typical peaks that indicate transitions between long-range ordered and paramagnetic phases. 
This suggests that emergent dynamics in the SkL phase  enhance $\lambda$. In both 6.4\% and 10.5\% Zn-substituted materials we see two peaks which coincide with the SkL regions identified above. 
To estimate the correlation time in the SkL phases, we compared these results and further transverse field $\mu$SR measurements on 6.4\% Zn-substituted Cu$_2$OSeO$_3$. From these\cite{yaouanc2011muon} we estimate the correlation time to range between approximately $\tau~=~\SI{10}{\nano\second}$ at \SI{48}{\kelvin} and $\tau~=~\SI{70}{\nano\second}$ at \SI{60}{\kelvin}.
Within the SkL phase the correlation time increases by almost a factor of two in the lower-temperature SkL phase, and approximately a third in the higher-temperature SkL phase.

Closer examination of the spectra measured for 6.4\% Zn substitution near the peaks in $\lambda$ suggests that they are better
described by the sum of two exponential functions with distinct relaxation rates $\lambda_{i}$,
This is seen in Fig.~\ref{fig:lambda}(a), where data plotted on a logarithmic scale shows  two approximately linear regimes (separated by a crossover at $\approx\SI{2.5}{\micro\second}$) with significantly different gradients, suggesting that  a two-exponential model provides a more accurate description of the data.
This strongly implies the occurrence of  two magnetically and spatially distinct classes of muon site, with different correlation times, since if there were only one muon site subject to fluctuations with two correlation times, the slower one would dominate, leading to a single relaxation rate~\cite{heffner2000observation}.
Moreover, we expect both classes of muon site to occur in the bulk of the material. 
For comparison, the same analysis was also carried out on LF-$\mu$SR data measured on a polycrystalline sample of pristine Cu$_2$OSeO$_3$ (where only one peak is seen). Here, a single exponential function models the data best, suggesting that there is only one magnetically distinct muon site in this material.
These measurements therefore provide unambiguous evidence for the coexistence of at least two magnetic phases in the bulk of polycrystalline Zn-substituted Cu$_{2}$OSeO$_{3}$.

Evidence for multiple phases in polycrystalline samples has come from structural and magnetic measurements. We may link these with
the observation of a correlation between
the lattice parameter and $T_{\mathrm{c}}$ for each phase in the polycrystalline samples. These are plotted in Fig.~\ref{Fig:max4}(a), labeled by nominal Zn-substitution. The fractional volume of each phase was also determined by both Rietveld refinement and magnetometry and  is shown in Fig.~\ref{Fig:max4}(b). There is good agreement between the two datasets. We therefore conclude that the polycrystalline Zn-substituted Cu$_{2}$OSeO$_{3}$ samples are inherently multiphase, featuring multiple, distinct Zn-doped Cu$_{2}$OSeO$_{3}$ phases. This then explains the origin of the split skyrmion pocket: each distinct phase exhibits the skyrmion spin structure at a different temperature, reflecting the differing values of $T_{\mathrm{c}}$. 

Although the estimated Zn content in our single crystal samples is, at most, 2.4\% and so we might not expect to resolve any splitting in the skyrmion phase, we note that in the 2.0\% substituted polycrystalline sample there is evidence for both structural and magnetic phase separation. We see no evidence for multiphase behaviour in our single crystals, suggesting that these are composed of a single phase with a single magnetic transition and should be expected to host a single skyrmion lattice phase. 

In conclusion, we have shown that the  splitting of the SkL phase observed exclusively in polycrystalline samples reflects the system's splitting into multiple Zn-substituted phases, each characterized by different magnetic behaviour. This demonstrates the importance of high-quality single crystal samples in the investigation of the magnetic properties of skyrmion hosting systems. 

This work is financially supported by EPSRC (EP/N032128/1).
Part of this work was performed at the ISIS Facility, Rutherford Appleton Laboratory,
 Diamond Light Source and the Swiss Muon Source, Paul Scherrer  Institut. We are grateful for 
the provision of beamtime. We thank Matthias Gutmann for useful discussions. Data will be made available via Durham Collections. 

%%%%%%%%%%%%%%%%%%%%%%%%%%%%%%%%%%%%%%%%%%%%%%%%%%%%%%%%%%%%%%%%%%%%%%%%%%%%%%%%%%%%%%%%%%%%%%%
%\bibliography{Ref}

\end{document}

\title{Supplemental material for Origin of skyrmion lattice phase splitting in Zn-substituted Cu$_{2}$OSeO$_{3}$}
\author{A. \v{S}tefan\v{c}i\v{c}}
\email{A.Stefancic@warwick.ac.uk}
\affiliation{University of Warwick, Department of Physics, Coventry, CV4 7AL, United Kingdom}
\author{S. Moody}
\affiliation{Durham University, Department of Physics, South Road, Durham, DH1 3LE, United Kingdom}
\author{T.J. Hicken}
\affiliation{Durham University, Department of Physics, South Road, Durham, DH1 3LE, United Kingdom}
\author{T.M. Birch}
\affiliation{Durham University, Department of Physics, South Road, Durham, DH1 3LE, United Kingdom}
\author{G. Balakrishnan}
\affiliation{University of Warwick, Department of Physics, Coventry, CV4 7AL, United Kingdom}
\author{S. Barnett}
\affiliation{Diamond Light Source, Harwell Science and Innovation Campus, Didcot OX11 0DE, United Kingdom}
\author{M. Crisanti}
\affiliation{University of Warwick, Department of Physics, Coventry,
  CV4 7AL, United Kingdom}
\affiliation{Institut Laue-Langevin,  Large Scale Structures Group,  71 avenue des Martyrs
CS 20156, 38042, Grenoble, Cedex 9,  France}
\author{J.S.O. Evans}
\affiliation{Department of Chemistry, Durham University, South Road, Durham, DH1 3LE, United Kingdom}
\author{S.J.R. Holt}
\affiliation{Durham University, Department of Physics, South Road, Durham, DH1 3LE, United Kingdom}
\author{K.J.A. Franke}
\author{P.D. Hatton}
\author{B.M. Huddart}
\affiliation{Durham University, Department of Physics, South Road,
  Durham, DH1 3LE, United Kingdom}
\affiliation{Durham University, Department of Physics, South Road, Durham, DH1 3LE, United Kingdom}
\author{M.R. Lees}
\affiliation{University of Warwick, Department of Physics, Coventry, CV4 7AL, United Kingdom}
\author{F.L. Pratt }
\affiliation{ISIS Facility, STFC Rutherford Appleton Laboratory, Chilton Didcot, Oxfordshire, OX11~0QX, United Kingdom}
\author{C.C. Tang}
\affiliation{Diamond Light Source, Harwell Science and Innovation Campus, Didcot OX11 0DE, United Kingdom}
\author{M.N. Wilson}
\affiliation{Durham University, Department of Physics, South Road, Durham, DH1 3LE, United Kingdom}
\author{F. Xiao}
\affiliation{Laboratory for Neutron Scattering, Paul Scherrer Institut, CH-5232 Villigen PSI, Switzerland}
\affiliation{Department of Chemistry and Biochemistry, University of Bern, CH-3012 Bern, Switzerland}
\author{T. Lancaster}
\email{tom.lancaster@durham.ac.uk}
\affiliation{Durham University, Department of Physics, South Road, Durham, DH1 3LE, United Kingdom}

\maketitle

\section{Sample preparation}

Polycrystalline $\text{Cu}_{2}\text{OSeO}_{3}$ materials with nominal Zn substitution values of $0,\ 2,\ 5,\ 8,\ 10, 12$ and $15\%$ were synthesised by thoroughly grinding together stoichiometric amounts of CuO ($99.99\%$, metals basis, Alfa Aesar), $\text{SeO}_2$ ($99.999\%$, trace metal basis, Acros Organics) and ZnO ($99.999\%$, Aldrich)  inside an argon-filled glove box.
The mixtures of powders were transferred into silica tubes, evacuated
and sealed. The samples were heated at a rate of $3.5~^{\circ}\text{C\ h}^{-1}$ to $650~^{\circ}\text{C}$, kept at this temperature for $96~\text{h}$, followed by water quench cooling.

\begin{figure}[t]
\includegraphics[width=\linewidth]{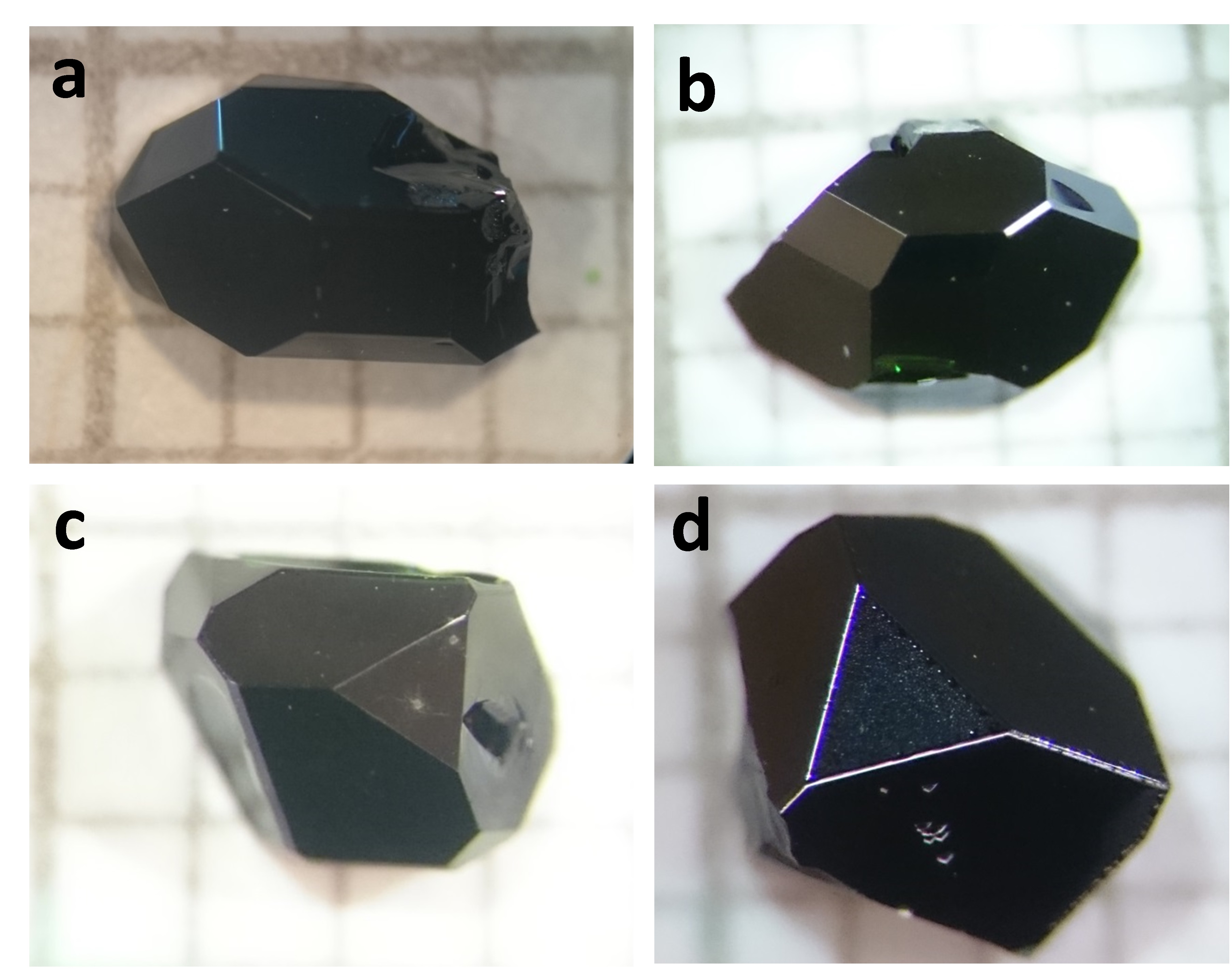}
\caption{Single crystals of $\text{Cu}_{2}\text{OSeO}_{3}$ with
(a) 0, (b) 0.5, (c) 1.8, and (d) 2.4\% Zn substitution levels respectively. Pictures were taken on millimetre paper.}
\label{Fig:Crystals}
\end{figure}

Single crystals of $\text{Cu}_{2}\text{OSeO}_{3}$ with nominal $0,\
2,\ 8$ and  $12 \%$ Zn substitution levels were grown by the chemical vapour phase transport (CVT) technique.
Polycrystalline materials $(2.5~\text{g})$, described above, and
$\approx 1.5$ to $2~\text{mg}/\text{cc}$ of the transporting agent, $\text{TeCl}_4$, were sealed in evacuated silica tubes. 
The growth of single crystals with Zn substitution levels between
$0$ and $2.4\%$ was achieved by heating the source part of the tube to
$640~^{\circ}\text{C}$ and the sink part to $550~^{\circ}\text{C}$ for four weeks.
Dark olive-green, almost black, single crystals of size $\approx 3 \times
3 \times 3~\text{mm}^3$ (see Fig.~\ref{Fig:Crystals}) were obtained. 

\section{X-ray measurements}
In-depth structural investigations 
of polycrstalline samples was carried out at Diamond Light Source
(Harwell Science \& Innovation Campus) using the high-resolution $3$-circle
diffractometer, I$11$ beamline, operating in transmission mode
equipped with $5$ multi-analysing crystals (MAC) detector. 
Data was collected at a wavelength of
$0.825251(10)$ \AA, 
in a $2\theta$ range between $0$ and $150^{\circ}$.
Rietveld refinements were carried out in TOPAS academic v$6.0$
software \cite{coelho2012topas}.  

The single crystal X-ray diffraction was performed on a 
a Fluid Film Devices 3-circle diffractometer using a Dectris Pilatus
2M detector at the I$19$ beamline, Diamond Light Source (Harwell Science \& Innovation Campus)\cite{cryst7110336}. 
Data collection at an energy of 17.9976~keV  was carried out on $20$ to $30~\mu$m size crystals,
cleaved from the single crystals shown in Fig.~\ref{Fig:Crystals}.

The xia2 \cite{xia2} and DIALS \cite{Winter:di5011} packages were used for data collection and
reduction, and absorption corrections were performed by numerical
integration using the SADABS software \cite{bruker2008inc}. 
Crystal structures were solved by a direct method and refined by
full-matrix least squares on F$^2$ for all data using SHELXT
\cite{sheldrick2015shelxt} and SHELXL \cite{sheldrick2015crystal}
embedded in Olex$^2$ \cite{dolomanov2009olex2} software.

The structural model of $\text{Cu}_{2}\text{OSeO}_{3}$
\cite{effenberger1986kristallstrukturen}, cubic $P2_1 3$ space group,
was refined to diffraction profiles of polycrystalline
samples. Results from the synchrotron measurements are given in Table~\ref{Tab:Rietveld}.
(We also performed laboratory XRD measurements and found that their resolution is insufficient to observe the splitting of the diffraction peaks.)
Selected bond lengths and bond angles for single crystal samples are collected in
Tables~\ref{tab:bondlengths} and \ref{tab:bondangles} respectively.

%%%%%%%%%%%%%%%%%%%% Table2: Powder X-Ray Refinement %%%%%%%%%%%%%%%%%%%%
\begin{table*}[b]
\centering
\begin{tabular}{c|cccccccc}
\hline \hline
\textbf{Sample} & $\boldsymbol{a}~\left(\textbf{\AA}\right)$ & $\boldsymbol{V}~\left(\textbf{\AA}^\textbf{3}\right)$ & \textbf{\% of }$\textbf{Cu}_\textbf{2}\textbf{OSeO}_\textbf{3}$ & \textbf{\% of }$\textbf{Cu}\textbf{O}$ & \textbf{\% of }$\textbf{Zn}_\textbf{2}\textbf{SiO}_\textbf{4}$ & $\boldsymbol{R}_\textbf{wp}$ & $\boldsymbol{R}_\textbf{exp}$ & $\boldsymbol{\chi}^\textbf{2}$ \\
\hline

0\% Zn & 8.923518(2) & 710.572(0) & 98.6(1) & 1.45(1) & 0.0         & 4.35 & 1.84 & 2.36 \\

\hline

2\% Zn & 8.924835(4) & 710.887(1)  & 83.1(1)   & 1.16(1) & 0.0         & 4.14 & 1.94 & 2.14 \\

       & 8.92718(4)  & 711.448(5)  & 11.8(1)  &
  &           &      &      &      \\

    & 8.92962(4)  & 712.033(9)  & 3.9(1)   &
  &           &      &      &      \\

\hline

4.1\% Zn & 8.925872(5) & 711.135(1) & 64.9(1)   & 1.3(1) & 0.9(1) & 3.98 & 1.87 & 2.13 \\

  & 8.92883(1)  & 711.842(3) & 26.5(1)   &   &           &      &      &      \\

  & 8.93135(3)  & 712.446(8) & 6.3(1)    &   &           &      &      &      \\

\hline

6.4\% Zn & 8.926116(9) & 711.193(2) & 31.4(1)   & 1.0(1)& 1.6(1) & 4.03 & 1.79 & 2.25 \\

  & 8.928525(7) & 711.769(2) & 53.2(1)   &   &           &      &      &      \\ 

  & 8.93142(1)  & 712.463(2) & 12.5(1)  &   &           &      &      &      \\

\hline

7.9\% Zn& 8.926530(8) & 711.292(2) & 34.1(1)   & 1.4(1) & 2.1(1) & 4.70 & 1.91 & 2.46 \\

  & 8.92928(1)  & 711.952(3) & 37.6(1)   &   &           &      &      &      \\

  & 8.931560(9) & 712.495(2) & 24.8(1)   &   &           &      &      &      \\

\hline

10.5\% Zn& 8.927459(5) & 711.514(1) & 36.5(1)   & 3.0(1) & 1.5(1) & 4.16 & 1.86 & 2.23 \\

  & 8.932180(3) & 712.644(1) & 59.0(1)  &   &           &      &      &      \\

\hline 

13.4\% Zn& 8.92731(1)  & 711.480(3) & 19.3(1)    & 4.9(1) & 1.6(1) & 3.96 & 1.93 & 2.05 \\

  & 8.931850(2) & 712.565(1) & 74.1(1)   &   &           &      &      &      \\

\hline \hline

\end{tabular}
\caption{Lattice parameters values, quantities of phases present in the polycrystalline samples of (Cu$_{1-x}$Zn$_x$)OSeO3 extracted from Rietveld refinements, along with qualities of fits.}
\label{Tab:Rietveld}
\end{table*}

\begin{table}
\begin{tabular}{c|c|c|c|c}
\hline\hline
Bonds & Pristine Cu2OSeO3 (\AA) & 2.4\% Zn (\AA) & uncertainty  (\AA)& difference (\AA)\\
\hline\hline
Cu$^{\mathrm{\parallel}}$ -- O1& 1.9251(10) &1.9246(15) &0.00180 &$- 0.0005$\\
Cu$^{\mathrm{\parallel}}$ -- O2& 1.9675(10) &1.9680(17) &0.00197 &$+ 0.0005$\\
Cu$^{\mathrm{\parallel}}$ -- O3& 2.0201(11) &2.0214(15) &0.00186 &$+ 0.0013$\\
Cu$^{\mathrm{\parallel}}$ -- O4 &1.9728(10) &1.9761(14) &0.00172 &$+ 0.0033$\\
Cu$^{\mathrm{\parallel}}$ -- O4'& 2.2847(10) &2.282(2) &0.00224 &$- 0.0027$\\
Cu$^{\mid}$ -- O1& 1.9225(10) &1.9265(15) &0.00180 &$+ 0.0040$\\
Cu$^{\mid}$ -- O2& 1.9077(10) &1.9095(15) &0.00180 &$+ 0.0018$\\
Cu$^{\mid}$ -- O3& 2.0851(11) &2.0829(16) &0.00194 &$- 0.0022$\\
\hline\hline
\end{tabular}
\caption{Bond lengths for single crystal samples, including overall
  uncertainty in the bond lengths. \label{tab:bondlengths}}
\end{table}

\begin{table}
\begin{tabular}{c|c|c|c|c|}
\hline\hline
Angles& Pristine Cu2OSeO3 (\AA) &
2.4\% Zn (\AA) &uncertainty  (\AA)& difference  (\AA)\\
\hline\hline
O1 -- Cu$^{\mathrm{\parallel}}$ -- O3&  79.28(4)$^{\circ}$& 79.28(6)$^{\circ}$& 0.07& $0.00$\\
O1 -- Cu$^{\mathrm{\parallel}}$ -- O4& 92.70(4)$^{\circ}$&  92.68(6)$^{\circ}$&  0.07& $- 0.02$\\
O2 -- Cu$^{\mathrm{\parallel}}$ -- O3 &100.59(4)$^{\circ}$& 100.68(6)$^{\circ}$& 0.07& $+ 0.09$\\
O2 -- Cu$^{\mathrm{\parallel}}$ -- O4 &87.76(4)$^{\circ}$& 87.68(6)$^{\circ}$& 0.07& $- 0.08$\\
O2 -- Cu$^{\mathrm{\parallel}}$ -- O4'& 79.55(4)$^{\circ}$& 79.67(6)$^{\circ}$& 0.07& $+ 0.12$\\
O1 -- Cu$^{\mid}$ -- O3& 77.74(4)$^{\circ}$& 77.72(6)$^{\circ}$& 0.07& $- 0.02$\\
O2 -- Cu$^{\mid}$ -- O3 &102.27(4)$^{\circ}$& 102.28(6)$^{\circ}$& 0.07& $+ 0.01$\\
O3 -- Cu$^{\mid}$ -- O3 &115.61(2)$^{\circ}$& 115.60(3)$^{\circ}$& 0.04& $- 0.01$\\
\hline\hline
\end{tabular}
\caption{Bond angles for single crystal samples, along with overall
  uncertainty in the bond angles. \label{tab:bondangles}}
\end{table}

\section{Magnetization}
\begin{figure}[t]
\includegraphics[width=\linewidth]{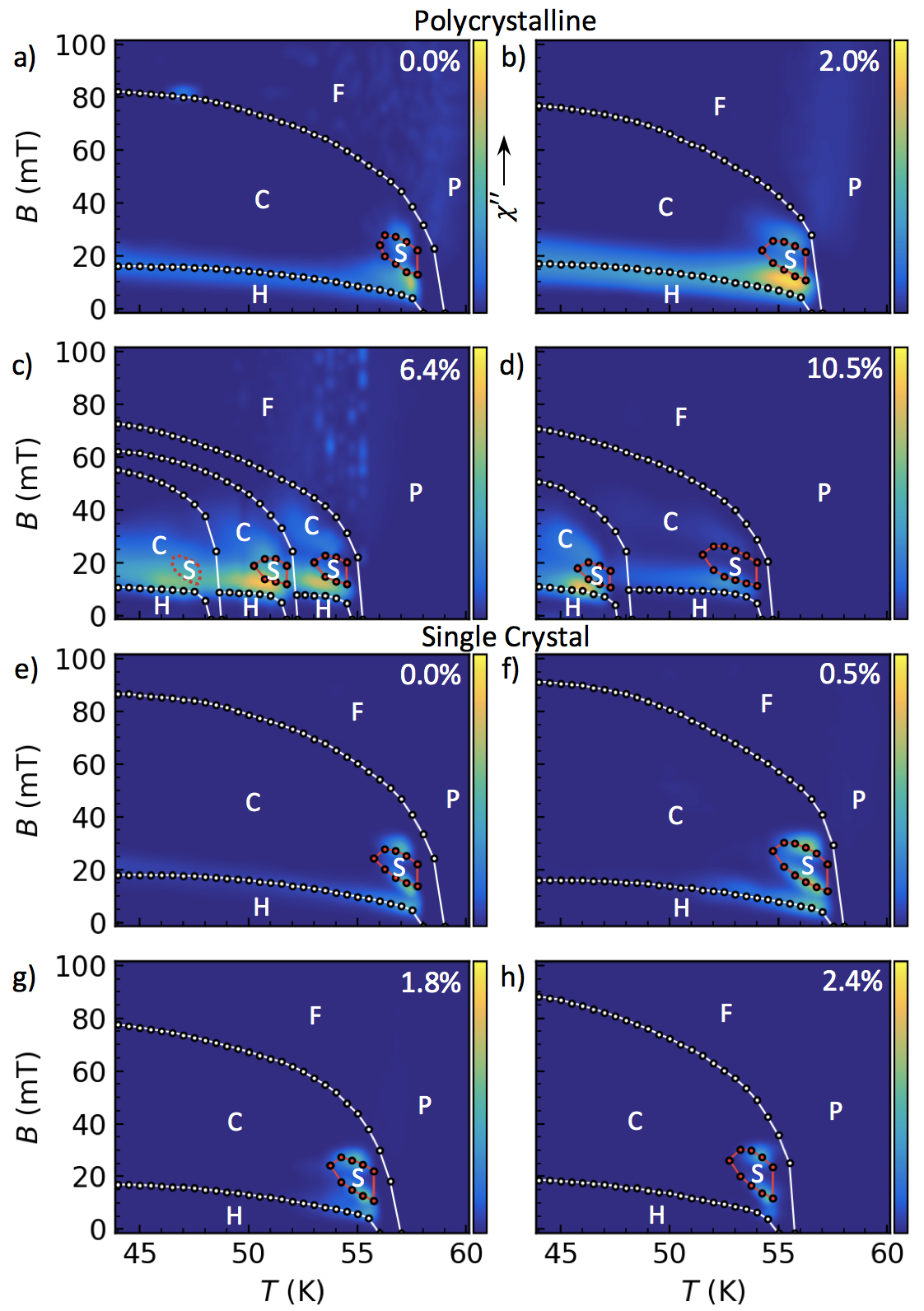}
\caption{Imaginary part $\chi''$ of the AC susceptibility for
  polycrystalline (left column) and single crystal (right colume) materials.}
\label{Fig:xipp_complete}
\end{figure}

A Quantum Design Magnetic Property Measurement System, MPMS-5S,
superconducting quantum interference device (SQUID) magnetometer was
used for investigation of the magnetic properties of the
polycrystalline samples and oriented single crystals as a function of
temperature and field. 
Temperature-dependent DC magnetic susceptibility ($\chi$) measurements
were carried out in an applied field of $250$~Oe in the temperature
region between $5$ and $110$ K under both zero-field-cooled (ZFC) and
field-cooled (FC) protocols.  
AC ($\chi'$ and $\chi''$) measurements in an AC driving field of 3~Oe
were measured at fixed temperature as a function of DC bias field between $0$ and $1000$ Oe 
in order to construct the field $(B)$ $vs$ temperature $(T)$ phase
diagrams. 
We assigned the phase boundary of the helical state and the conical state by a sharp increase in the AC susceptibility. The signature of the skyrmion state in these measurements was a sharp drop in the AC susceptibility within the conical phase.
Single crystals used for the AC measurements were aligned with $B$
$\mid\mid$ to $[111]$,  with alignment achieved using a Photonic
Science X-ray Laue back-scattering imaging system. 
Ordering temperatures found from the derivatives of the magnetization are given in Tables~\ref{Tab:Magnet} and
\ref{Tab:Phase_Trans_both}. 
The
imaginary part $\chi''$ of the AC susceptibility is shown for 
  polycrystalline and single crystal materials in
  Fig.~\ref{Fig:xipp_complete}. 

\begin{table*}[b]
\centering
\begin{tabular}{c|ccccccc}
\hline \hline
\textbf{Sample}	&\textbf{0 \%}	&\textbf{2.9 \%}	&\textbf{4.1 \%} & \textbf{6.4 \%}&\textbf{7.9 \%}&\textbf{10.5 \%}&\textbf{13.4 \%}\\
\hline
$T_{\mathrm{c}}$ (K)	& 58.62(1)	& 51.6(2)	& 49.2(2)	& 48.7(1)	& 48.39(7)	& 47.97(3)	& 47.89(3)	\\
				&			& 53.7(3)	& 52.10(8)	& 52.32(6)	& 51.61(7)	& 54.80(8)	& 54.3(1)	\\
				&			& 57.33(2)	& 55.99(2)	& 55.45(5)	& 55.21(5)	&			&			\\
		\hline
\textbf{Weight}	& 1.0		& 0.07(2)	& 0.090(9)	& 0.16(1) 	& 0.26(1)	& 0.64(1)	& 0.80(1)	\\
				&			& 0.06(2)	& 0.235(9)	& 0.43(1)	& 0.36(1)	& 0.35(2)	& 0.20(2)	\\
				&			& 0.870(7)	& 0.675(8)	& 0.41(1)	& 0.38(1)	&			&			\\
\hline \hline

\end{tabular}
\renewcommand{\baselinestretch}{1.00}
\caption{Ordering temperatures and the quantities of phases present in
  polycrystalline  samples extracted from the gradient of $M-T$ magnetometry.} 
\renewcommand{\baselinestretch}{1.15}
\label{Tab:Magnet}
\end{table*}

\begin{table*}[t]
\centering
\begin{tabular}{c|c|c|c|c}
\hline \hline
\textbf{Substitution levels} & \textbf{0\% Zn} & \textbf{0.5\% Zn} & \textbf{1.8\% Zn} & \textbf{2.4\% Zn}\\
\hline
$T_\text{c}~(\text{K})$  in single crystals& 58.30(2)&58.09(2)&56.96(2)&55.98(2)\\
\hline \hline
\end{tabular}
\caption{Ordering temperatures $(T_\text{c})$ observed in
  single crystals; obtained from $M-T$ magnetization.}
\label{Tab:Phase_Trans_both}
\end{table*}

\section{Muon-spin relaxation}

In a muon-spin relaxation ($\mu$SR) experiment,\cite{blundell1999spin} spin-polarized positive muons are implanted into the sample and subsequently decay into a positron, emitted preferentially in the direction of the muon's instantaneous spin vector. The measured quantity is the time-dependent positron asymmetry $A(t)$, which is proportional to the muon ensemble's spin polarization.
Longitudinal field (LF) $\mu$SR measurements (where an external field
$B_{0}$ is applied parallel to the initial muon-spin direction) were
carried out on polycrystalline samples of Cu$_2$OSeO$_3$ of 0\%, 6.4\%
and 10.5\%  Zn-substitution on the HiFi instrument at the ISIS Neutron and Muon Source, Rutherford Appleton Laboratory, UK.
Transverse-field (TF) $\mu$SR measurements (applied field  perpendicular to the initial spin direction) were made on the 6.4\% Zn-substituted Cu$_2$OSeO$_3$ using the GPS instrument at the Swiss Muon Source (S$\mu$S), Paul Scherrer Institut, Switzerland.
In all cases the samples were packed in silver foil envelopes (foil thickness \SI{12.5}{\micro\meter}) and mounted on a silver plate.
For all measurements, samples were zero-field cooled to the minimum temperature, before the field was applied and measurements were made on warming. 

\begin{figure}[t]
\includegraphics[width=\linewidth]{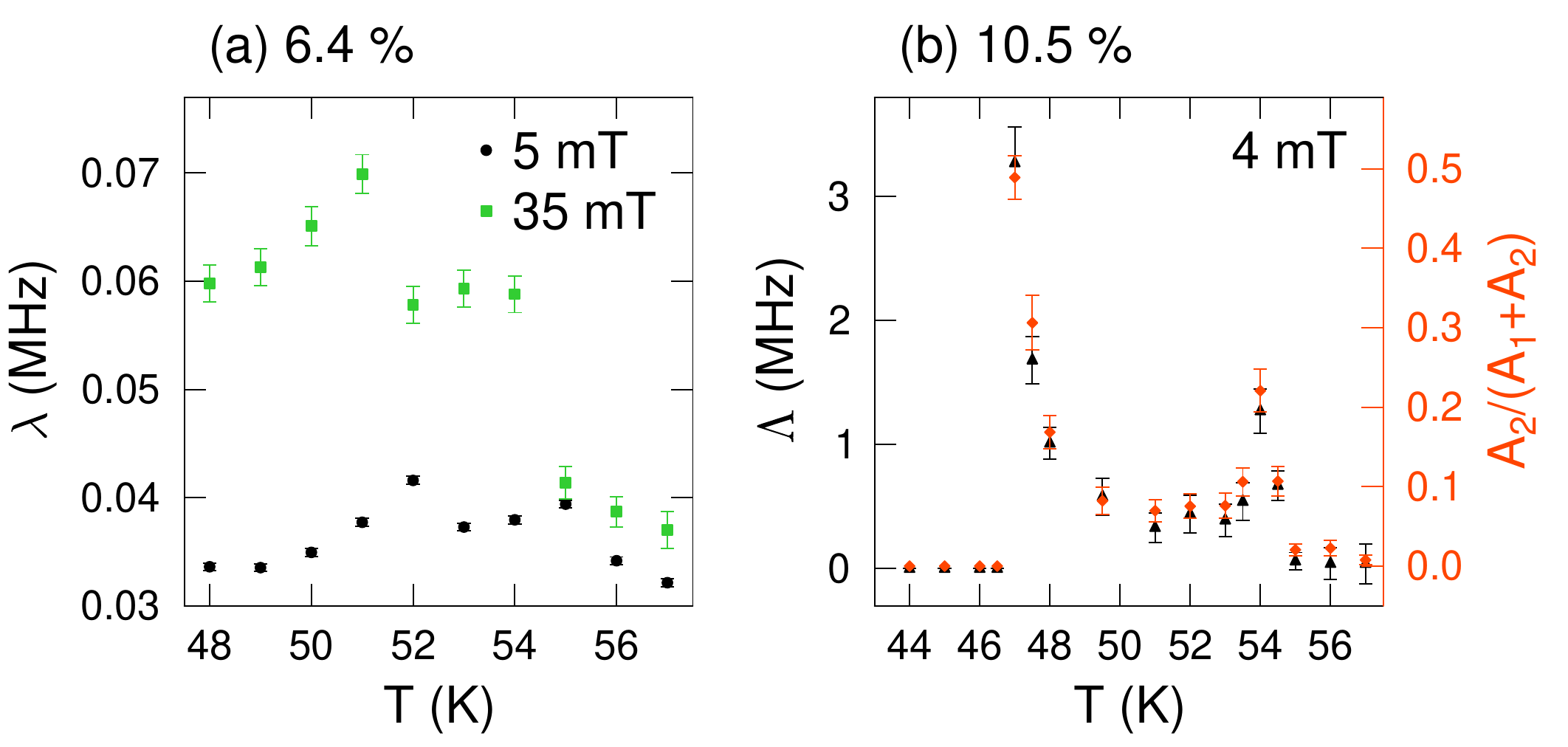}
\caption{(a) Relaxation rates for 6.4\% Zn-doped Cu$_2$OSeO$_3$ at two
  different applied fields, extracted using fits to
  Eqn.~\ref{eqn:single_exp_fit}. (b) Transverse relaxation rate
  $\Lambda$ and amplitude of oscillating component (both from
  Eqn.~\ref{eqn:4mT_12pc_fit}) for 10.5\% Zn-doped Cu$_2$OSeO$_3$ in
  an applied field of \SI{4}{\milli\tesla}. 
\label{fig:lambdanotskyrmion}}
\end{figure}

In addition to the  LF-$\mu$SR measurements described in the main text, measurements were made at fields outside the SkL phases. 
For 6.4\% Zn-doped Cu$_2$OSeO$_3$ these fields were $B_{0}~=~\SI{5}{\milli\tesla}$ (below the minimum field at which the SkL is realized) and \SI{35}{\milli\tesla} (above the maximum field at which the SkL is realized), whereas for 10.5\% Zn-doped Cu$_2$OSeO$_3$ a field of \SI{4}{\milli\tesla} was applied.
Spectra for 6.4\% Zn-doped Cu$_2$OSeO$_3$ were found to relax monotonically and were fitted with a single exponential decay
\begin{equation}\label{eqn:single_exp_fit}
A\left(t\right) = A_1e^{-\lambda t} + A_\text{bg},
\end{equation}
where the non-relaxing background $A_\text{bg}$ results from spins that do not relax over the timescale of the measurement, such as those that implant within the Ag foil or sample holder.
The extracted values of the longitudinal relaxation rate $\lambda$ are shown in Fig.~\ref{fig:lambdanotskyrmion}(a).
Note that these peaks appear much sharper than the peaks when cutting through the SkL, as seen in the main text.
This shows that cutting through the skyrmion lattice causes an enhanced $\lambda$ when compared to just cutting through the phase transition.

In the 10.5\% Zn-doped Cu$_2$OSeO$_3$ data an oscillation was observed for $T>\SI{47}{\kelvin}$  (due to spins precessing in the sum of the applied and internal fields within the ISIS time window).
To account for this, these data were fitted to the function
\begin{equation}\label{eqn:4mT_12pc_fit}
A(t) = A_1e^{-\lambda_1t} + A_2e^{-\Lambda t}\cos(\gamma_{\mu}Bt + \phi) + A_\text{bg} ,
\end{equation}
where $B$ was kept fixed at \SI{4}{\milli\tesla} for all temperatures where the oscillation was resolvable, and $\phi$ is a phase offset.
Fig.~\ref{fig:lambdanotskyrmion}(b) shows transverse relaxation rate $\Lambda$ and oscillation amplitude $A_2$ (normalised by the total relaxation).
The results of these fits (Fig.~\ref{fig:lambdanotskyrmion}(b)) show peaks in relaxation rates and amplitudes indicative of phase transitions (although we note that they occur at temperatures slightly below those  identified via AC susceptibility).
This provides evidence for two phase transitions, supporting the claim
that there are (at least) two different Zn-doping levels in the
powder, and each of these has a different transition temperature. 

%\bibliography{Ref}